# 140 Characters to Victory?: Using Twitter to Predict the UK 2015 General Election

Pete Burnap[*,1], Rachel Gibson[2], Luke Sloan[3], Rosalynd Southern[2] and Matthew Williams[3]

[1] Cardiff School of Computer Science & Informatics, Cardiff University
[2] Cathie Marsh Institute for Social Research, University of Manchester
[3] Cardiff School of Social Sciences, Cardiff University

**Introduction**

The election forecasting 'industry' is a growing one, both in the volume of scholars producing forecasts and methodological diversity. In recent years a new approach has emerged that relies on social media and particularly Twitter data to predict election outcomes. While some studies have shown the method to hold a surprising degree of accuracy there has been criticism over the lack of consistency and clarity in the methods used, along with inevitable problems of population bias. In this paper we set out a 'baseline' model for using Twitter as an election forecasting tool that we then apply to the UK 2015 General Election. The paper builds on existing literature by extending the use of Twitter as a forecasting tool to the UK context and identifying its limitations, particularly with regard to its application in a multi-party environment with geographic concentration of power for minor parties.

**Using Twitter to Predict Elections: The Story So Far**

The increasing use of social media globally has dramatically increased the amount of data available to track and predict trends in the economy, public opinion and population health (Pries et al., 2013; Ortiz et al., 2011; Mellon, 2014). The use of Twitter data to forecast elections has become increasingly prominent since the start of the current decade, however, there is no consensus over how to forecast with Twitter and the findings to date have been mixed. Tumasjan et al's (2010) study of the 2009 German Federal election constitutes the first published attempt to use Twitter to estimate a national election result. As with the studies that have followed, it was not a genuine forecast in that it was conducted post-election. The results were encouraging, however, in that the authors claimed a high degree of accuracy for their analysis which compared the share of mentions of the six most prominent parties and associated politicians in tweets over a five week period prior to election day, to their final vote share. Criticism of the study and its somewhat crude 'more tweets, equals more votes' premise soon followed. In particular Jungherr et al. (2012) noted the lack of methodological justification for the time period used and how the tweets were captured. Re-running the analysis over a longer time span that ran closer to the election day resulted in a higher mean adjusted error (MAE) of 2.13 compared to the 1.65 of the original study and a much higher MAE than traditional polls.

Complementing this specific rebuttal, Gayo-Avello (2011, 2012) identified more general problems in the use of Twitter to predict election outcomes. Topmost among his concerns was the need to produce a true forecast, i.e. one that was issued prior to the election. In addition he stressed the need to take into account the biases within the Twitter using population and existing power distribution among the candidates and parties being studied. Finally he called on analysts to incorporate Tweet sentiment into the computation rather than rely simply on volume. Other published and unpublished empirical studies produced around the same time raised some major questions about the accuracy of twitter as a forecasting tool (Gayo-Avello et al. 2011; Bermingham and Smeaton 2011; O'Connor et al., 2010; Metaxas et al., 2011; Skoric et al. 2012; Sang and Bos, 2012)

---

[*] burnapp@cardiff.ac.uk

Subsequent studies appear to have taken on board some of Gayo-Avello's advice with some more encouraging results. DiGrazia et al. (2013) for example added a range of individual and district level controls to a regression model using Twitter mentions to predict vote share for candidates in the 2010 and 2012 U.S. Congressional elections. The authors concluded a positive and statistically significant relationship remained even after accounting for incumbency and parties' existing levels of popularity. Franch (2013) took a more dynamic approach and examined sentiment expressed toward the three main party leaders across a number of social media platforms, including Twitter in the lead-up to the 2010 UK election. Using an auto regressive integrated moving average (ARIMA) model he regressed daily measures of party support from Yougov polls on their social media popularity scores and generated a set of final predictions that were within 1 percent of the three parties actual vote share. Ceron et al. (2014) used sentiment analysis to compute a Twitter popularity rating for Italian political leaders in the 2011 parliamentary elections and candidates in the French 2012 Presidential election. According to the authors the results were almost analogous to the predictions based on polls and in line with academic forecasts using offline data (Nadeau et al. 2012). Finally Caldarelli at al. (2014) introduced a 'relative support' parameter to their analysis that produced an 'instant indicator' of the comparative strength of two parties on Twitter (using mentions). This was used to predict the election results for the four main parties in the 2013 Italian parliamentary elections. While the results confirmed to the authors that Twitter is 'an effective way to get indications of election outcomes' they admitted that it over-predicted the vote of the two main parties. The error, they argued, followed from the inclusion of the party leaders' names as search terms (Monti and Berlusconi), both of whom were former Prime Ministers and the latter was on trial at the time.

Overall, therefore, the extant literature appears to offer grounds for expecting that Twitter can serve as a useful tool in predicting electoral outcomes across a variety of national contexts, subject to certain corrective steps. Our approach is start with a 'baseline' model of Twitter forecasting that borrows from the KISS principle used in Agent Based Modelling whereby one starts with the most basic and transparent model which can then be built upon (Axelrod, 1997). In doing so we follow three of Gayo-Avello's main recommendations. First we offer a genuine forecast based on Twitter data harvested no later than one month prior to election day. Second we adjust our forecast to take into account the sentiment of the tweet. Finally in calculating our predictions on seat rather than vote share we take into account the existing distribution of parliamentary representation and party power within each constituency.

**Methodology**
We began by collecting data from the Twitter streaming API (Burnap et al. 2014). Tweets were selected if they included party and/or leader names, as shown in Table 1. The search was not case sensitive so it effectively collected mentions with upper and lower case spelling. The collection was commenced on the 28th November 2014 and contained 13,899,073 tweets by the time the forecast was calculated on 9th March 2015. See Appendix Table 1 for search terms.

After harvesting the Twitter sample we then applied automated sentiment analysis using software developed by Thelwall et. al (2010), which allocates a string of text a positive and negative score ranging from -5 (extreme negative) to +5 (extreme positive), where each score is produced based on words in the string that are known to carry such emotive meaning (e.g. 'love'=5, hate='-4'). Where a tweet contained more than one of the search terms (e.g. "I'm voting *Labour* because I can't stand *David Cameron*"), we removed the tweet from the sample to avoid misallocating the positivity in the tweet. Clearly, in the example, the positivity is directed towards Labour, but the automatic identification of sentiment direction was beyond the scope of this study.

We first calculated sentiment scores for each tweet and produced a list of all tweets with associated positive and negative sentiment scores. Applying a rationale that positive tweets containing party or leader names can be treated as vote intentions, we removed all tweets where sentiment scores were below -1, and kept those between -1 and +5. The value of the remaining sentiment scores were summed to produce a *party sentiment score* and *a leader sentiment score*. Scores for leaders representing the same party (e.g. Natalie Bennett and Caroline Lucas) were combined, as were party mentions (e.g. Tories, Conservatives etc). The reason for summing all tweet sentiment scores as opposed to counting the number of positive mentions was to record the overall magnitude of the sentiment. In a situation where Labour had the same number of positive tweets as the Conservatives, the summed sentiment score would differentiate the parties where the average sentiment was higher for one than the other. The summed sentiment scores for all parties and their leaders (e.g. Tories, Conservatives, David Cameron etc.) were then combined to produce a single *positive party sentiment sum* for each party. All positive party sentiment sums were combined to calculate the total sentiment, which was used to normalise the positive party sentiment sum for each party, with respect to all other parties, thus producing a party-specific Twitter positive sentiment proportion (see Table 2).

Visual inspection of the data identified an unusually high level of false-positives for the search terms "Labour" and "Greens", due to the different contexts in which these terms can be used. Using 3-way human annotation, where three individuals manually annotated a random sample of 1,000 tweets including these terms according to whether each tweet was actually related to the UK Labour or Green Parties, we identified that 78.9% of tweets containing the word "Labour" were actually about the Labour Party and only 19.4% of the tweets containing the term "Greens" were actually about the Green Party. The reason for the high proportion of type one errors associated with the Green Party was due to the bulk of activity focusing on the Australian Green Party. This weighting was applied when calculating positive Twitter proportions and had the effect of reducing the overall representation of these party mentions in the relative proportions. Table 1 reports our estimates of vote shares.

In a final step we converted our vote shares into a seat forecast (see Table 2). To do so we applied our vote share to the UK 2010 results and calculated a measure of national swing which was then applied on a constituency by constituency basis to produce an estimate of which party would win a given seat. For example, in Halesowen and Rowley Regis, West Midlands, the vote share in 2010 was CON=41.2, LAB=36.6, LIB=14.8 (CON WIN). Using the Twitter vote share to calculate change from 2010, the projected split becomes CON=34.4, LAB=35.9, LIB-3.8 (LAB WIN), showing a swing from Conservative to Labour. This process was performed for all seats in the UK, and the final number of seats won was calculated for each party by selecting the maximum value for each seat (see Table 3).

**Discussion and Conclusions**
The results show the likely outcome is a hung parliament with the Labour party gaining most seats. While the predictions for the other parties look to be within an expected range there is clearly a significant under-estimation of SNP seats. This result points to the limitations of using Twitter to forecast in multi-party systems where there is a 'majority' regionalist party. Without a means of geo-locating tweets there will always be an under-estimation of such support since the assumption of our calculations is that individuals are randomly distributed across the UK. Given the reduction in N after geocoding methods are applied (i.e. tweets including a precise location) – only around 1% of tweets are retained – we opted to retain our larger sample with an acceptance of the dilution in SNP support.

More generally we consider our analysis presents a step forward in the literature in that our model meets at least three of the core criteria set out in the literature to reach a minimal acceptable standard for forecasting. Future applications need to incorporate methods for geo-location and also to apply corrections for bias in sample demographics. Fortunately work is currently underway by the authors (Sloan et al. 2013, Sloan et al. 2015) to address these issues and is predicted to be available in time for the next UK General Election.

**Table 1 – Twitter sentiment and vote share (some minor parties removed where we did not collect the leader)**

|  | Party +ve Sentiment Share | Leader +ve Sentiment | Total +ve Sentiment | +ve Twitter Proportion |
|---|---|---|---|---|
| **Conservative** | 0.107393085 | 0.404418497 | 0.511811582 | 0.292351853 |
| **Labour** | 0.305422798 | 0.189374473 | 0.494797271 | 0.28263311 |
| **Lib Dem** | 0.014556207 | 0.063081068 | 0.077637275 | 0.044347182 |
| **SNP** | 0.10048713 | 0.059566354 | 0.160053484 | 0.091424138 |
| **Green** | 0.048597728 | 0.124554097 | 0.044856119 | 0.02562226 |
| **UKIP** | 0.276503519 | 0.138688431 | 0.41519195 | 0.237161761 |
| **DUP** | 0.02433284 | 0.004705975 | 0.029038815 | 0.016587259 |
| **Plaid** | 0.001658883 | 0.001890266 | 0.003549149 | 0.002027309 |
| **Sinn Fein** | 1.34E-05 | 0.013720838 | 0.013734232 | 0.007845129 |

**Table 2 – Vote share change from 2010 based on Twitter and Projected Seat Wins**

|  | 2010 Share | Twitter Share | Change | Seats |
|---|---|---|---|---|
| Conservative | 36.1 | 29.3 | -6.8 | 285 |
| Labour | 29.0 | 28.3 | -0.7 | 306 |
| Lib Dem | 23.0 | 4.4 | -18.6 | 21 |
| SNP | 1.7 | 9.2 | +7.5 | 9 |
| Green | 1 | 2.3 | +1.3 | 1 |
| UKIP | 3.1 | 23.8 | +20.7 | 5 |
| DUP | - | 1.7 |  |  |
| Plaid | 0.6 | 0.2 | -0.4 | 3 |
| Sinn Fein | - | 0.8 |  |  |

APPENDIX
Twitter search terms

| Leaders | Parties |
| --- | --- |
| David Cameron | Conservative Party |
| Ed Miliband | Tory Party |
| Nick Clegg | Tories |
| Nigel Farage | Labour Party |
| Natalie Bennett | UKIP |
| Caroline Lucas | Green Party |
| leanne wood | Labour |
| nicola sturgeon | SNP |
| alex Salmond | Scottish National Party |
| peter Robinson | lib dems |
| gerry adams | liberal democrats |
| | greens |
| | UK independence party |
| | BNP |
| | British National Party |
| | DUP |
| | Democratic Unionist Party |
| | Sinn Fein |
| | SDLP |
| | UUP |
| | Ulster Unionist Party |
| | Plaid |